# EFFICIENT BLIND SEARCH: OPTIMAL POWER OF DETECTION UNDER COMPUTATIONAL COST CONSTRAINTS

By Nicolai Meinshausen, Peter Bickel and John Rice

*University of Oxford and University of California, Berkeley*

Some astronomy projects require a blind search through a vast number of hypotheses to detect objects of interest. The number of hypotheses to test can be in the billions. A naive blind search over every single hypothesis would be far too costly computationally. We propose a hierarchical scheme for blind search, using various "resolution" levels. At lower resolution levels, "regions" of interest in the search space are singled out with a low computational cost. These regions are refined at intermediate resolution levels and only the most promising candidates are finally tested at the original fine resolution. The optimal search strategy is found by dynamic programming. We demonstrate the procedure for pulsar search from satellite gamma-ray observations and show that the power of the naive blind search can almost be matched with the hierarchical scheme while reducing the computational burden by more than three orders of magnitude.

**1. Introduction.** "What is the most efficient way to look for a needle in a large haystack?" This is, in essence, a problem one faces in some astronomy projects. Huge amounts of data are gathered. These could be luminosities from thousands of stars [Alcock et al. (2003)], photon arrival times from potential pulsars [Kanbach et al. (1989), Gehrels and Michelson (1999)] or measurements to detect gravitational waves [Abramovici et al. (1992)]. Some of the encountered problems might be more accurately formulated as follows: "What is the most efficient way to look for a needle in several thousand haystacks?" Take the search for gamma-ray pulsars as an example. For each of many observed sources, photon arrival times are recorded and the question is whether there is any periodicity in the arrival times of the photons. To this end, the power-spectrum has to be calculated on a very fine grid of frequencies. The number of frequencies to be searched scales proportionally to the length $T$ of the observations. Additionally, rotation-powered pulsars

"spin down" over time, losing some of their rotational energy. Searching over frequency alone is not sufficient and the drift has to be accounted for. If the drift is linear over the time period of observations, the number of drift values to be searched (to guarantee some chosen power of detection) scales as $T$. The number of hypotheses thus increases like $T^3$ and is already unbearably high for the Energetic Gamma Ray Experiment Telescope (EGRET) on the *Compton Gamma Ray Observatory* satellite observations [Kanbach et al. (1989)], with roughly 100 billion frequency-drift combinations to be searched for each source to determine whether there is any periodicity in the photon arrival times. Thus, in a blind search for radio-quiet pulsars among the EGRET sources, Chandler et al. (2001) used a 512 processor supercomputer. The upcoming *Gamma-ray Large Area Space Telescope* (*GLAST*) survey will be faced with even higher computational challenges as observational times are longer.

Some related problems in astronomy include the detection of gravitational waves and the search for variable stars. To verify the existence of gravitational waves, the *Laser Interferometer Gravitational-wave Observatory* (*LIGO*) is in operation [Abramovici et al. (1992)] and the *Laser Interferometer Space Observatory* (*LISA*) is proposed [Danzmann et al. (1996)]. The precision of the experiment is unprecedented. Any search for gravitational waves will have to be at a very fine grid in frequency space, posing computational challenges. Likewise for the search for stars with periodically changing luminosity. These fluctuations can sometimes indicate the presence of a planet and some exoplanets have been found in this way. Surveys with the potential for the search for stars with variable luminosity include the *Taiwan-America Occultation Survey* (*TAOS*) [Alcock et al. (2003)], the *Panoramic Survey Telescope And Rapid Response System* (*Pan-STARRS*) [Kaiser et al. (2002)] and the *Large Synoptic Survey Telescope* (*LSST*) [Tyson (2002)]. We will mostly be taking the search for gamma-ray pulsars as an example, though, using data from the EGRET mission [Kanbach et al. (1989)], data being from the 3rd EGRET catalogue [Hartman et al. (1999)].

*Resolution levels and hierarchical search.*    The proposed hierarchical search scheme rests on a simple underlying idea. Instead of testing every single of, potentially, billions of hypotheses individually, some kind of "aggregate" or "summary" statistic is calculated for sub-groups of hypotheses. These could, in the simplest case, be the average test statistics over each sub-group. The needle in the haystack corresponds to a single or a few true nonnull hypothesis (the needle) to be detected among, potentially, billions of true null hypotheses (the haystack). If the evidence at the individual level is strong enough for the true nonnull hypotheses to stand out from billions of null hypotheses, it is certainly still high enough to produce "suspiciously" high values of any sensible aggregate statistics (like an average) over sub-groups of, say, tens or hundreds of hypotheses. One could first calculate the aggregate statistics over all sub-groups. In a second stage, one can then examine the promising candidate groups in more detail. The computational savings consist of not spending too much effort on uninteresting sub-groups of hypotheses. Such a two-stage procedure begs the question of optimality. Maybe a three-stage procedure is even better, doing a pre-selection of promising sub-groups at a very coarse level before entering the two-stage procedure? Or maybe four or five-stage procedures? Also, it is unclear what the optimal size of sub-groups is. Should one aggregate over ten hypotheses or thousands? We propose a fitting algorithm that can, in principle, answer these questions. Instead of asking the traditional statistical question,

> "What is the most powerful test procedure under a constraint on the false-positive rate?"

we rather ask

> "What is the most powerful test procedure under a constraint on (a) the false-positive rate and (b) the computational cost of the procedure?"

By taking the constraint on the computational resources into account, the problem becomes much richer and challenging. We only try to illuminate some aspects of taking this step and think that problems of this type will play some role beyond the astronomical problems ("astronomical" both in scope and size) mentioned above.

*Related work.* Such ideas go back at least to proposals for group testing [Dorfman (1943)] and work on sequential estimation [Kiefer and Sacks (1963), Bickel and Yahav (1967)], and more recently have been considered in the physics community. A simple hierarchical setup, yet only in a two-layer version, was proposed in Brady and Creighton (2000). Later, Cutler, Gholami and Krishnan (2005) showed that the computational costs could be reduced even further by adding a third stage and considered also more general N-stage searches. In their regime, looking for gravitational-wave pulsars, a three-stage turned out to be optimal. We try to formulate the problem so that it is applicable to a broader array of problems than detection of gravitational waves. Our main contribution is the formulation of the optimal solution in terms of dynamic programming (DP). That way, all involved constants, number of layers, resolution levels and other involved parameters can be chosen automatically in a nearly optimal way. Moreover, we show how DP can be implemented in a computationally efficient manner for a hierarchical tree.

A related issue occurs also for fast online pattern recognition. In a widely cited paper, Viola and Jones (2001) proposed "boosting cascades" as a very fast approach for object recognition in images. In each "cascade" step, a computationally cheap algorithm (in effect, a few steps of a boosting algorithm) reduces the number of potential targets by a large factor, without losing real objects of interest on the way. A finer distinction has then to be made only for a greatly reduced number of potential objects. In similar spirit, Blanchard and Geman (2005) examine the optimality of hierarchical testing designs, using the related assumption of tests with zero type II error (not losing any object of interest on the way). They derive interesting results about optimality conditions for any hierarchical testing scheme under these assumptions. We take a more pragmatic approach and try to derive an optimal detection scheme under nonnegligible type II errors.

**2. Hierarchical search.** After introducing some notation, the example of pulsar detection is described in more detail and the notion of hierarchies and resolution levels is illustrated at hand of this example.

The notation for the "naive" search is introduced. In the naive search, each hypothesis out of a potentially very large number $n$ of hypotheses is tested. The parameter of interest is denoted by $\theta_j$, $j = 1, \ldots, n$, for each individual hypothesis. The associated null hypotheses are

$$H_{0,j} : \theta_j \in \Theta_0, \qquad j = 1, \ldots, n,$$

$$H_{A,j} : \theta_j \in \Theta_A, \qquad j = 1, \ldots, n.$$

Associated with each test $j = 1, \ldots, n$ is a real-valued test statistic $Z_j$. Assume for simplicity that each test statistic $Z_j$ follows, marginally and conditional on $\theta_j$, the same distribution for all $j = 1, \ldots, n$. Each null hypothesis $H_{0,j}$ can be rejected if $Z_j > q$, where $q$ is chosen such that, for true null hypotheses, the probability of exceeding the threshold is limited by $P(Z_j > q) \leq \alpha$. The level $\alpha$ can be chosen to achieve any desired family-wise error rate or false discovery rate. For control of the family-wise error rate, one can, for example, use a simple Bonferroni correction. We will not be concerned with the choice of the threshold $q$ in the following and simply assume it to be fixed.

Each test incurs a certain computational cost and it might be too expensive to do all $n$ tests. The question is then if we can find procedures that maintain good power while being computationally cheaper than the naive search over each individual hypothesis.

2.1. *Example: pulsar detection.* Using satellite gamma-ray detectors, the EGRET project recorded photon arrival times from a number of known and unknown sources. For the (known) pulsar PSR B1706–44, for example, 1072 photons with energies above 200 MeV were observed over a time-span of $T = 1205197$ seconds, roughly 14 days. These pulsars performed more than $10^8$ rotations in this observational time, meaning that the chance to see a photon in any given cycle is very low. What is a sensible statistic to detect periodicity in these arrival times? The problem is studied in Bickel, Kleijn and Rice (2007a), Bickel, Kleijn and Rice (2007b). Surprisingly and fortunately, even though the detection sensitivity of the detector changes over time (e.g., because the earth moves between the satellite and the object of interest), the distribution of the test statistic does not depend on it. To test for periodicity at a given frequency $\omega$ and drift $\dot{\omega}$ with photon arrival times $t_1, \ldots, t_m$, we use in the following simply the power in the first harmonic,

$$(1) \qquad F(\omega, \dot{\omega}) = \frac{2}{m} \left| \sum_{j=1}^{m} \exp(2\pi i \phi_j) \right|^2 \qquad \text{where } \phi_j = \omega t_j + \dot{\omega} t_j^2 / 2.$$

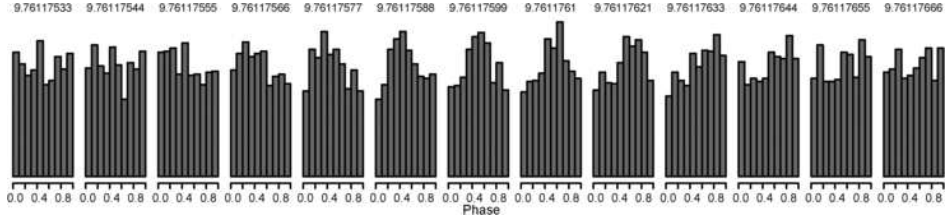

FIG. 1. *Histogram of the phase of photon arrival times for pulsar PSR B1706–44. The probed frequency in Hz is shown on the top of each histogram and is increasing slightly from left to right. The true frequency is in the center and shows the clearest deviation from uniform arrival times. The drift was adjusted to the true drift of this pulsar. In a blind search, the frequency range to be searched ranges from about 1 Hz to 40 Hz.*

If $\phi(T) \gg 1$, the null distribution of the statistic is approximately chi-squared with two degrees of freedom. This is the Rayleigh test statistic for periodicity [Mardia and Jupp (1999)]. Related test statistics, for example, statistics that use higher harmonics, are discussed in Bickel, Kleijn and Rice (2007a), Bickel, Kleijn and Rice (2007b). It would be straightforward to use such an alternative statistic in the hierarchical search. For simplicity of exposition, we will use (1). As alluded to above, the test statistic would have to be calculated over a fine grid for possible frequency-drift pairs $(\omega, \dot{\omega})$ to have a reasonable chance of detection. Figure 1 illustrates the effect of even a slight misspecification of the frequency. The drift has to be matched even better, as the number of gridpoints necessary scales linearly in $T$ for frequency. This precision is needed in order that an arrival time at the beginning of the record be in phase with one at the end. For the same reason, the scaling of the number of necessary grid values is $T^2$ for drift, as any mismatch $\delta\dot{\omega}$ in drift leads to a phase shift $\delta\dot{\omega} T^2 / 2$ at the end of the observational period.

*Tapering and blocked search.* The function $F$ can be "smoothed out" by so-called "blocking" of the observational time. Instead of calculating the power in each frequency-drift combination over the entire length of the observations, one can divide the observational time into $2^\kappa$ blocks $B_1, \ldots, B_{2^\kappa}$ of equal length, with some value of $\kappa \in \mathbb{N}$. Calculating (1) separately for each block and summing the result over all blocks, one obtains the modified statistic

$$(2) \qquad F^\kappa(\omega, \dot\omega) = \frac{2}{m} \sum_{k=1}^{2^\kappa} \left| \sum_{j \,:\, t_j \in B_k} \exp(2\pi i \phi_j) \right|^2.$$

In essence, the phase information between successive blocks is lost. The resulting statistic has not as sharp a peak for the true frequency-drift pair as $F$. It is, on the other hand, less sensitive to misspecifications of these values. The number of grid points to be searched is of order $T/2^\kappa$ for frequency and of order $T^2/2^{2\kappa}$ for drift.

The "smoothing out" effect of blocking can be observed in Figure 2. For pulsar PSR B1706–44, the function $F^\kappa$ is shown for the values $\kappa = 0, \ldots, 4$ in the vicinity of the "true" frequency-drift of pulsar PSR B1706–44. It becomes evident that sensitivity to the drift parameter decreases more rapidly for an increasing number of blocks, compared with the sensitivity to frequency. This supports the previous observation that the number of gridpoints needs to scale like $T$ for frequency but like $T^2$ for drift. The hierarchical search navigates its way from the coarsest level to the finest resolution level, trying to detect all peaks while minimizing the computational cost.

2.2. *Tree structure.* The basic idea of a hierarchical search is to start at a coarse resolution level, where a small computational cost is sufficient to divide the search space into "interesting" regions (which warrant further exploration at finer resolution levels) and "uninteresting" regions (on which no further computational resources should be spent).

The hierarchical setup can most conveniently be represented by trees. Each tree has $G$ layers, or generations, of nodes, corresponding to the various resolution levels. The coarsest resolution corresponds to layer $\ell = 1$, while the finest resolution level, where final test decisions are made, corresponds to level $\ell = G$.

At each layer, some observations can be made. The number of potential observations is increasing for finer resolution levels. Figure 3 illustrates the basic setup. For simplicity, assume that the number of original tests can be written as $n = 2^D$. The first layer contains $2^{D-G+1}$ measurements

$$X_1^{(1)}, \ldots, X_{2^{D-G+1}}^{(1)}.$$

Each successive layer contains then twice as many potential measurements as the previous layer, up to the original $2^D$ measurements for the final layer.

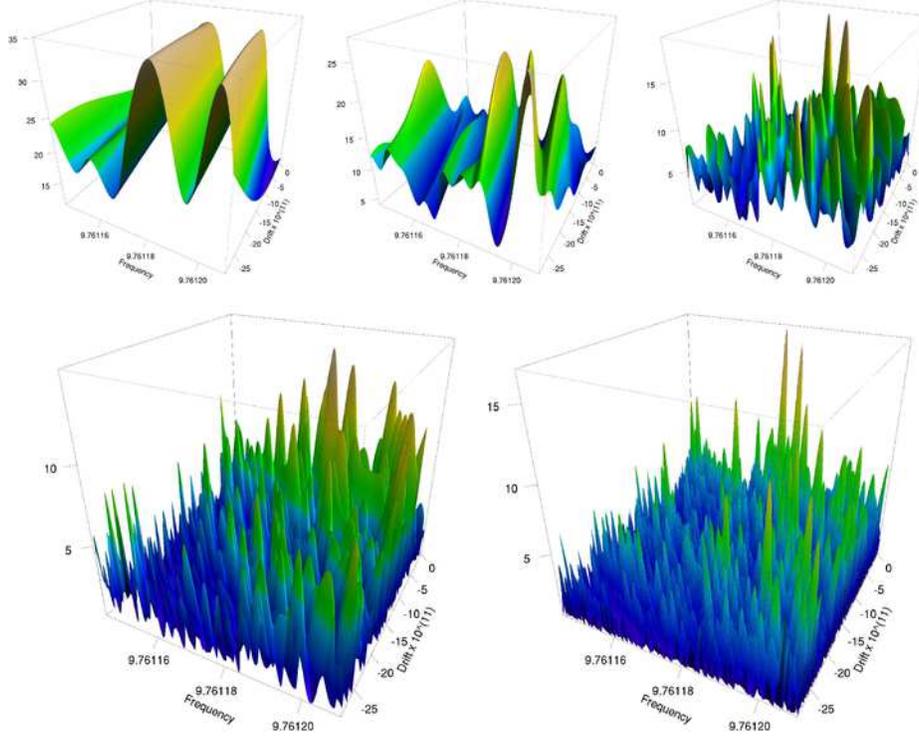

FIG. 2. *The frequency-drift plot for pulsar PSR B1706–44, showing the power in the first harmonic at different "resolution" levels. From top left to bottom right, the number of blocks decreases from 16 over 8, 4 and 2 to 1 at the finest resolution. The true frequency and drift of the pulsar is seen as a sharp peak at the finest resolution at the bottom right. The peak is "smeared out" at lower resolution levels.*

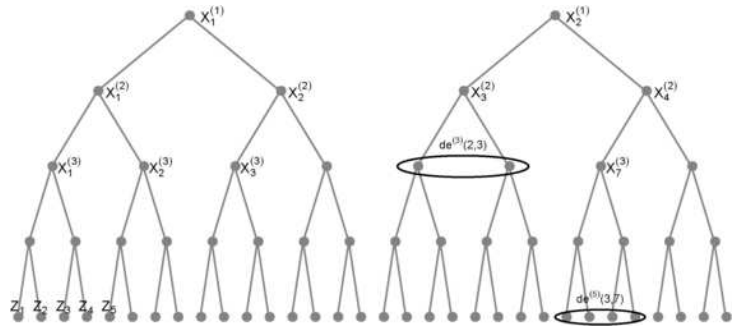

FIG. 3. *Illustration of the hierarchical setup. The ellipsoids indicate that, for example, the descendants of node 7 in layer 3 can be referred to as* $\mathrm{de}^{(5)}(3,7)$, *if one is looking at descendants in the fifth layer. The original test statistics* $Z_1, \ldots, Z_n$ *form the lowest layer of the hierarchy and are identical to* $X_1^{(5)}, \ldots, X_n^{(5)}$.

The layers are organized in a binary tree. In each layer, the nodes $v = 1, \ldots, 2^{D-G+\ell}$ of the tree correspond to the potential measurements that can be made. Each node in layers $\ell$ up to $G-1$ has two children in the next layer $\ell + 1$. The search proceeds along these tree branches. If the observation at a node is "promising," the children are examined in more detail. Or, maybe, some further descendants are examined directly. If the observation at node $v$ is not "promising" enough, search in this subtree is stopped.

Take a node $v$ in layer $\ell$. All descendants of this node in layer $s > \ell$ are referred to as de$^{(s)}(\ell, v)$. Figure 3 shows two examples. The final layer $\ell = G$ contains all the original $n$ observations $Z_1, \ldots, Z_n$, where each $Z_v$ is identical to $X_v^{(G)}$, and we will mainly use the latter notation in the following.

We have formulated the problem in terms of binary splits. The approach can easily be generalized to form $K$-ary splits. However, the case of binary splits is quite general. The search strategy will not necessarily have to observe nodes in every layer and can "jump" across layers. Hence, there is no negative effect if the binary splits cause more layers to be present than necessary for an efficient search, as the optimal search strategy will simply ignore these superfluous layers. The number $G$ of layers has to be chosen a priori. Again, there is in general no harm in introducing too many layers, as the search strategy will simply ignore layers that are not needed for efficient search.

2.3. *Hierarchical search strategies.* Any value in the tree can only be observed upon "payment" of a computational price $C_\ell$, which depends, potentially, on the layer $\ell$ of the tree. The question is then: if one has already observed some nodes, which ones should be observed next? And which subtrees are unpromising candidates? The descendants of those unpromising candidates should then be discarded in a further search.

If the observed value of a node is sufficiently "interesting" (typically, sufficiently high), the search proceeds to the next layer in the subtree, that is, formed by taking the observed node as a root node. Or the search could skip the next layer entirely and proceed with the second next layer and so on. On the other hand, if the observed value indicates that this region of the search space does not warrant further investigation, the search can stop completely in the entire subtree.

Let $O$ be a vector with the same dimensionality as $X$. The value of $O_v^{(\ell)}$ at node $v$ in layer $\ell$ is given by

$$O_v^{(\ell)} = \begin{cases} 1, & \text{if the value } X_v^{(\ell)} \text{ has been observed,} \\ 0, & \text{if the value } X_v^{(\ell)} \text{ has not (yet) been observed.} \end{cases}$$

The search strategy $S$ can be formulated separately for each layer. The strategy $S$ can thus be written as a vector $(S^{(1)}, \ldots, S^{(G)})$. Upon observing

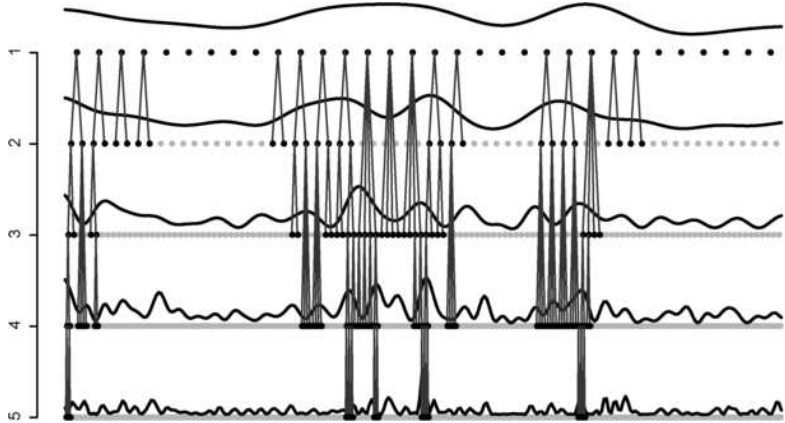

FIG. 4. *The outcome of the hierarchical search for pulsar PSR B1706–44 in the vicinity of the "true" frequency (in the center of the plot) and for the "true" drift. At the lowest resolution in level 1, all frequencies are examined on a coarse grid. The line above each layer indicates the value of the test statistic at each node (where a node corresponds to a frequency). All observed test statistics are shown in black, while unobserved frequencies are gray. Promising candidate frequencies are then searched on a finer grid in layer 2. For sufficiently high values, layer 2 is even skipped and the search continues directly on a finer grid in layer 3. The peak in the lowest layer around the true frequency is found. The total number of tests made is 176, compared to 512 frequencies that would have to be searched if only the layer at the finest resolution would be available. A lot of activity in this picture is triggered by the peak at the "true" frequency (and some smaller peaks next to the main peak as a result); for other, uninteresting, regions with no peak (which make up the vast majority of a typical blind search), the proportion of tests made is more than three orders of magnitude smaller than the number of potential tests in the final layer.*

the value $X_v^{(\ell)}$ in layers $\ell = 1, \ldots, G$, the strategy $S^{(\ell)}$ is a mapping

$$S^{(\ell)} : \mathbb{R} \mapsto \{0, \ell + 1, \ldots, G\},$$

saying in which generation to search next. The strategy depends in this formulation only on the layer $\ell$ and not the individual node $v$. This eases the computational burden when fitting the optimal strategy and the optimal strategy is of this form if all $X_v^{(\ell)}$ in the same layer $\ell$ are exchangeable. A value of 0 means that no further search takes place in the subtree formed by taking node $v$ as root node. A value $s > \ell$ means that generation $s$ is searched next. This generation $s$ can be larger than $\ell + 1$. In this case, one or more generations are skipped and the search continues directly on a much finer grid. Continuing the search after the final layer is not sensible and we use the convention that $S^{(G)}(x) = 0$ for all values $x$ in layer $G$.

The hierarchical search proceeds then as follows:

1. **Initialize.** For the layer $\ell = 1$, set $O_v^{(1)} = 1$ for all $v$ (so that all values in the layer are observed). Initialize $O_v^{(\ell)} = 0$ for all $v$ in all other layers $\ell = 2, \ldots, G$.
2. **Loop.** Do the following successively for all layers $\ell = 1, \ldots, G$. For each node $v$ in layer $\ell$,
   (a) Calculate the value $x = X_v^{(\ell)}$ if and only if $O_v^{(\ell)} = 1$.
   (b) Let $s = S^{(\ell)}(x)$. If $s \neq 0$, set $O_u^{(s)} = 1$ for all descendants $u \in \mathrm{de}^s(\ell, v)$ of node $v$ in layer $s$.

The algorithm above shows the implementation for the search strategy under the assumption that the strategy, which consists of $G$ functions $S^{(1)}, \ldots, S^{(G)}$ is known. See Figure 4 for an example. To find an "optimal" strategy, one first needs to formulate the objectives. Here, we are trying to minimize the computational cost of the search while maximizing the chance of detecting truly interesting events (false null hypotheses in the final layer).

2.3.1. *Computational cost.* The computational cost of a search is simply the sum of the computational costs of observing nodes. A node $v$ in layer $\ell$ has been observed if $O_v^\ell = 1$, according to the notation introduced above. As the cost of observing any node in layer $\ell$ is $C_\ell$, the total cost is

$$\sum_{v,\ell} C_\ell 1\{O_v^{(\ell)} = 1\}.$$

The computational cost of a strategy $S$ is the expected value of the quantity above,

$$(3) \qquad \mathrm{Cost}(S) := E\left(\sum_{v,\ell} C_\ell 1\{O_v^{(\ell)} = 1\}\right).$$

The goal is to choose the strategy $S$ in a way to keep the computational cost as small as possible while maximizing power.

Note that we work with the idealized assumption that the computational cost of observing $N$ nodes in a given layer of the tree is linear in $N$. In practice, this assumption often does not hold up. For our example of pulsar detection, a fast fourier transform (FFT) can be employed to search over $N$ nodes at a computational cost of $N \log N$. Specifically, a single application of FFT can compute the test statistic in frequency drift space $(\omega, \dot\omega)$ along lines of constant ratio $\omega/\dot\omega$ [Ransom, Eikenberry and Middleditch (2002)]. There are potentially several ways of incorporating FFT into the proposed hierarchical search for pulsar detection. The hierarchical search could, for example, decide which lines of constant ratio $\omega/\dot\omega$ to examine. Details of this and other schemes are beyond the scope of the present manuscript, which merely tries to give a broadly applicable framework for a hierarchical search.

2.3.2. *Power.* The power of a strategy $S$ is determined by the probability of being able to reject false null hypotheses. Possible candidates for rejection are all nodes $v$ in the final layer $G$ whose value $X_v^{(G)}$ surpasses the critical threshold $q$. The crucial additional requirement is now that the value of those nodes has been observed. The power of the testing scheme is measured by

$$(4) \qquad \mathrm{Power}(S) := E\left(\sum_v 1\{X_v^{(G)} \geq q\} 1\{O_v^{(G)} = 1\}\right).$$

The power is clearly maximal if all leaf nodes are observed and is then equal to the power achieved by a naive search.

As an alternative to the definition above, one might want to count only rejections of those nodes which correspond to false null hypotheses. Rejecting true null hypotheses is clearly not desirable. We chose the definition above as it allows to quantify the "success" of the search scheme in situations where no knowledge is available as to which hypotheses would be truly false or true null hypotheses. More fundamentally, the search strategy is just a first step, narrowing down the number of hypotheses to be tested in the final layer. All these tested hypotheses can only be rejected if the corresponding values of the test statistics are above the critical value $q$. The overall procedure will have the same power as a computationally expensive naive blind search (where all hypotheses are tested) if we can observe all nodes in the final layer, where the test statistic is above $q$. With (4), one measures thus simply the additional loss of possible rejections resulting from the fact that not all hypotheses in the final layer are observed.

2.3.3. *Optimal strategies.* There is a tradeoff between the computational cost of a search strategy and its power. At one extreme, power could be maximized by searching over all hypotheses. This naive search strategy attains maximal power but increases the computational burden to unbearable levels. At the other extreme, cost could be minimized by dropping all searches immediately, with the result of zero cost yet also zero detection power. Clearly, neither of these two extremes is desirable.

For a given affordable computational cost $c$, an "optimal" strategy attains the maximal power under the computational constraint,

$$\arg\max_S \mathrm{Power}(S), \qquad \text{subject to } \mathrm{Cost}(S) \leq c.$$

It will be more convenient to formulate the problem in a different yet equivalent way. For each cost constraint $c$, there is a value $\lambda$ such that the optimal strategy is given by $S_\lambda$, defined as

$$(5) \qquad S_\lambda = \arg\max_S \{\mathrm{Power}(S) - \lambda \mathrm{Cost}(S)\}.$$

In most cases, one would typically be interested in a whole family of solutions, for different values of $\lambda$. This allows the value of $\lambda$ to be chosen once the entire tradeoff between power and cost is known. The question arises whether the optimal strategy $S_\lambda$ can be computed for any given value of the tradeoff parameter.

Before embarking on the details of the hierarchical search scheme, we would like to remark that the proposed strategy optimizes some aspects of the hierarchical search but certainly not all. For example, the observations in individual nodes are values of the test statistics $F^\kappa$, defined in (2). These test statistics are a means of smoothing out the detection statistic, and there are other ways. One is proposed in Bickel, Kleijn and Rice (2007b). The detection statistic is analytically integrated over frequency bands and this is accomplished by an expansion in eigenfunctions. The hierarchical scheme would then depend at each level on the widths of those bands, and also perhaps on the number of terms to retain in the expansions. A choice also has to be made for the total number $G$ of layers in the hierarchy. As remarked above, there is in general no harm in introducing too many layers $G$, as the search strategy will simply ignore superfluous layers. The optimal values for other parameters of the algorithm, like the choice of the test statistic, could be found by incorporating them into the dynamic programming framework.

2.4. *Dynamic programming.* To obtain a close approximation to the optimal solution $S_\lambda$, one can use dynamic programming [Bertsekas (1995)]. The optimal solution can be found by working down the hierarchy. The first nontrivial task is the strategy for layer $G - 1$. It can be found without knowledge about the optimal strategies for layers below $G - 1$. Once $S^{(G-1)}$ is known, strategy $S^{(G-2)}$ can be found, using the previously found approximately optimal strategy for layer $G - 1$. Proceeding in this way, the optimal strategy can be computed for every layer.

2.4.1. *Payoff.* The "payoff" $P$ is a random real-valued function for each node in the tree. For node $v$ in layer $\ell$, the payoff measures the "success" in the subtree, that is, induced by taking node $v$ in layer $\ell$ as root node. For a given realization, the payoff is then the number of correctly rejected null hypotheses in this subtree less the spent computational cost, weighted by $\lambda$, under a given strategy $S$.

$$
P_{\lambda,v,S}^{(\ell)} = \sum_{u \in \mathrm{de}^{(G)}(\ell,v)} 1\{X_u^{(G)} > q\} 1\{O_u^{(G)} = 1\}
$$
(6)
$$
- \lambda \sum_{\ell^* = \ell+1}^{G} \sum_{u \in \mathrm{de}^{(\ell^*)}(\ell,v)} C_{\ell^*} 1\{O_u^{(\ell^*)} = 1\}.
$$

The first part is the number of rejected hypotheses among all descendants in the final layer $G$. The second part measures the computational cost spent in the subtree. The sum stretches over all layers $\ell^*$ with $\ell^* > \ell$ and is by definition 0 if $\ell = G$. Note that the payoff is a function of the strategy $S$, as the strategy influences the observations $O$.

Using the definition of the payoff above, it becomes clear that the optimal strategy $S_\lambda$ maximizes the expected payoff at the first layer,

$$S_\lambda = \arg\max_S E\left(\sum_v P^{(1)}_{\lambda,v,S}\right).$$

This definition is equivalent to (5). The payoff function defined above will be useful to characterize the optimal strategy across all layers of the tree.

While the optimal strategy maximizes the expected payoff, one can conversely define a "value function" to characterize the payoff under such an optimal strategy.

2.4.2. *The value function $V$.* The value function $V$ at node $v$ in layer $\ell < G$ measures, conditional on the observation $X^{(\ell)}_v$, the expected payoff under the best possible strategy,

$$(7) \qquad V^{(\ell)}_{\lambda,v}(x) = \max_S E(P^{(\ell)}_{\lambda,v,S} \mid X^{(\ell)}_v = x).$$

The value function depends on the tradeoff parameter $\lambda$ and the layer $\ell$ under consideration. For the most part, we will be considering distributions of $X$ for which the value function above does not depend on the specific node $v$ in the layer. This holds true if the distribution of $X$ is identical under an exchange of nodes in the same layer of a tree (which includes exchange of the relevant subtrees attached to these nodes). We will write simply $V^{(\ell)}_\lambda$. Instead of looking at the value function directly, it is useful to look at the "continuation value," which measures the expected value if taking a specific action in the next step and following the optimal strategy thereafter.

2.4.3. *The continuation value $Q$.* The value function is calculated under the assumption that the optimal strategy is known completely. The "continuation value," in contrast, assumes only that the optimal strategy is known *after* the next necessary action has been taken; for details and more motivation, see Bertsekas (1995).

At node $v$ in layer $\ell < G$, the set of actions $s$ that can be taken is either to drop all further search (which corresponds to action "$s = 0$") or to proceed in layer $\ell^*$ with $\ell < \ell^* \le G$ (which corresponds to action "$s = \ell^*$"). The continuation value for some node $v$ in layer $\ell$, when taking action $s$, is given by

$$(8) \qquad Q^{(\ell)}_{\lambda,v,s}(x) = \sum_{u \in \mathrm{de}^{(s)}(\ell,v)} \{E(V^{(s)}_{\lambda,u} \mid X^{(\ell)}_v = x) - \lambda C_s\}.$$

The descendants $de^{(s)}(\ell, v)$ are, by convention, the empty set if $s = 0$ and the continuation value is in this case 0 (as it should be as all search is terminated). Again, we will mostly be working with situations where the value function does not depend on the specific node in a given layer and the continuation value can then be written as

$$(9) \qquad Q_{\lambda,s}^{(\ell)}(x) = \begin{cases} 0, & \text{if } s = 0, \\ 2^{s-\ell}(E(V_\lambda^{(s)} \mid X^{(\ell)} = x) - \lambda C_s), & \text{if } s \neq 0. \end{cases}$$

For the decision to drop all further search ($s = 0$), the continuation value is clearly zero, as no costs will be incurred but no detections will be made either. For $s \neq 0$, the set of descendants is of size $2^{s-l}$ (as we are working with binary trees).

The connection between the value function and the continuation value is simple. The value function (7) is the continuation value under the best decision,

$$(10) \qquad V_\lambda^{(\ell)}(x) = \max_{s \in \{0, \ell+1, \ldots, G\}} Q_{\lambda,s}^{(\ell)}(x).$$

The best decision is the one that leads to the highest continuation value

$$(11) \qquad S_\lambda^{(\ell)}(x) = \operatorname*{arg\,max}_{s \in \{0, \ell+1, \ldots, G\}} Q_{\lambda,s}^{(\ell)}(x).$$

If the continuation values are known, the value function follows immediately by (10). Also, the optimal action follows by (11). It is thus sufficient to fit the continuation values.

2.4.4. *Dynamic programming.* As motivated above, to find the optimal strategy, it is sufficient to fit the continuation values. Starting at the lowest layer, the fit of all possible $G - \ell$ continuation values in layer $\ell$ requires only the previously fitted continuation values of layers $\ell'$ with $\ell' > \ell$. Starting at the leaf nodes in the lowest layer, the value function can be computed as follows:

(1) **Initialize leaf nodes.** Set current generation $\ell$ to the generation $G$. The value function is

$$V_\lambda^{(G)}(x) = 1\{x \geq q\}.$$

(2) **Compute continuation value.** Move one generation down, $\ell \leftarrow \ell - 1$. The continuation value $Q_{\lambda,s}^{(\ell)}(x)$ is given for actions $s \in \{\ell+1, \ldots, G\}$ by

$$(12) \qquad Q_{\lambda,s}^{(\ell)}(x) = 2^{s-\ell}(E(V_\lambda^{(s)} \mid X^{(\ell)} = x) - \lambda C_s).$$

For $s = 0$, the continuation value vanishes identically.

(3) **Compute value function.** The value function for generation $\ell$ is given by

$$V_\lambda^{(\ell)}(x) = \max_{s \in \{0, \ell+1, \dots, G\}} Q_{\lambda,s}^{(\ell)}(x).$$

(4) **Loop.** If $\ell > 1$, go to Step 2. Else stop.

The most challenging step is the computation of the continuation value in (12), as it involves computing the expectation of the value function, conditional on the observations in lower layers. The approach we choose here is Monte Carlo (MC) sampling, fitting the continuation values by a nonparametric regression.

2.5. *Computing the optimal strategy by simulation.* Solving the dynamic programming problem explicitly is not feasible in general and one has to resort to simulation to approximate the conditional expectation in (12). Even a MC approach has to be well designed, due to the potentially very high number of tests involved and the enormous sizes of the trees involved. In situations we are interested in, the number of nodes in a tree can be in the billions. It is then infeasible to hold anything in memory which scales linearly with tree size. Only small subsets of the tree can be held in memory at any given time.

2.5.1. *Simulating "paths."* It might be very costly to simulate a single realization of $X$ over the entire tree. Moreover, to do so would be inefficient, as it would yield many simulations toward the leaves but just a limited number of samples for higher layers closer to the root nodes.

An alternative approach is to simulate $X$ for random "paths" in the tree. Let $(v_1, \dots, v_G)$ be a random path in the tree (where node $v_\ell$ sits in layer $\ell$ for all $\ell = 1, \dots, G$) in the sense that:

- if $\ell > 1$, node $v_\ell$ has to be in the set $\mathrm{de}^\ell(\ell - 1, v_{\ell-1})$ of descendants of the node $v_{\ell-1}$;
- the density is uniform over all such paths.

Consider that one would draw $m$ samples of $X$ over the entire tree. At the same time, $m$ random paths are drawn. The realization of $X$ along the corresponding path $(v_1, \dots, v_G)$ is then denoted for all $i = 1, \dots, m$ by

(13) $$X^{(1),i}, \dots, X^{(G),i}.$$

Specifying the layer $\ell$ is sufficient, as in each layer, only a single node (viz., $v_\ell$) is observed.

2.5.2. *Path payoff.* Definition (6) of "payoff" requires that the realization of $X$ over the entire tree is known. As we are (by computational necessity) working only with paths, we define the "path payoff" as the best estimate of the "payoff" under this constraint. The basic restriction is that one observes only one node $v$ in each generation. One can correct for that fact by multiplying every rejection and all the computational costs along the path by an appropriate factor. The path payoff $\hat{P}$ is defined by

$$(14) \quad \hat{P}^{(\ell)}_{\lambda,v,S} = 2^{G-\ell} 1\{X^{(G)}_{v_G} > q\} 1\{O^{(G)}_{v_G} = 1\} - \lambda \sum_{\ell^*=\ell+1}^{G} 2^{\ell-\ell^*} C_{\ell^*} 1\{O^{\ell^*}_{v_{\ell^*}} = 1\}.$$

At each layer, only one node is observed (the node in the path) and the payoff at this node, multiplied by an appropriate factor, is an unbiased estimate of the payoff across all nodes in this layer. The path payoff is an unbiased estimate of the payoff (6). If we let $X$ be the vector which contains all observations from the tree, then

$$(15) \qquad\qquad E(\hat{P}^{(\ell)}_{\lambda,v,S} \mid X) = P^{(\ell)}_{\lambda,v,S},$$

where the expectation is with respect to randomly sampled paths. Note that, conditional on $X$, the payoff is a constant and the path payoff depends only on the randomly chosen path. As the value function $V^{(\ell)}_\lambda(x)$ is the expected payoff, one obtains an unbiased estimate of the value function through the path payoff. For a node $v$ in generation $\ell$,

$$V^{(\ell)}_\lambda(x) = E(P^{(\ell)}_{\lambda,v,S} \mid X^{(\ell)}_v = x).$$

Thus, using (15),

$$(16) \quad V^{(\ell)}_\lambda(x) = E(E(\hat{P}^{(\ell)}_{\lambda,v,S} \mid X) \mid X^{(\ell)}_v = x) = E(\hat{P}^{(\ell)}_{\lambda,v,S} \mid X^{(\ell)}_v = x),$$

where $S$ is the optimal strategy. Samples from the path payoff are thus unbiased estimates of the value function and can be used to fit both the value function and the continuation values. The path payoff under the optimal strategy $S$ is clearly unknown, as the optimal strategy is unknown. Yet one can replace the optimal strategy in the computation of $P^{(\ell)}_{\lambda,v,S}$ above by an approximately optimal strategy.

2.5.3. *Implementing dynamic programming.* As mentioned above, the key to the implementation of dynamic programming is that the path payoff in layer $\ell < G$ is only a function of the continuation values in layers $\ell + 1, \ldots, G$.

Given a set of Monte Carlo samples as in (13), the goal is to fit all the continuation value functions as defined in (12). Starting at $\ell = G - 1$, the fitted continuation value is used to compute the path payoffs for the layer above. These path payoffs are, in turn, again the basis for fitting the continuation values in this layer. The procedure works then up the layers, as described in Section 2.4.4. The major step is fitting the continuation values in (12).

Suppose one fits the continuation value $Q_{\lambda,s}^{(\ell)}$ for layer $\ell$ and action $s \in \{0, \ell+1, \ldots, G\}$. Using the definition from (13), the value of $X$ in the $i$th MC sample in layer $\ell$ (in which there is only one node $v_\ell$ in the realized random path) is denoted by $X^{(\ell),i}$. Let $X^{(\ell),MC}$ be the $m$-dimensional vector that contains the observations in the $\ell$th layer from all $m$ paths,

$$X^{(\ell),MC} = (X^{(\ell),1}, \ldots, X^{(\ell),m}).$$

Let $\hat{P}_{\lambda,s}^{(\ell),i}$ be the path payoff for the $i$th MC sample in layer $\ell$ (where the node is determined by the random path), when choosing action $s \in \{0, \ell+1, \ldots, G\}$. In analogy to above, denote by $P^{(\ell),MC}$ the vector of all payoffs across all paths,

$$(17) \qquad \hat{P}_{\lambda,s}^{(\ell),MC} = (\hat{P}_{\lambda,s}^{(\ell),1}, \ldots, P_{\lambda,s}^{(\ell),m}).$$

The path payoff in layer $\ell$ depends on the strategy $S$ in layers $\ell+1, \ldots, G$, and one sets $S$ to be the strategy, that is, determined by the already fitted continuation values for these layers. Leaving off the subscript $s$, the value $P_\lambda^{(\ell)}$ is the payoff when choosing the optimal decision at layer $\ell$. The value $\hat{P}_\lambda^{(\ell),i}$ is the payoff for the $i$th MC sample when choosing the decision $s$ according to the fitted continuation values in layer $\ell$ itself.

By combining the computation of the continuation value (12) with the equivalence relation (16) between the value function and the path payoff, it follows that, for all $i = 1, \ldots, m$ and $s \in \{\ell+1, \ldots, G\}$,

$$Q_{\lambda,s}^{(\ell)}(x) = 2^{s-\ell}\{E(P_\lambda^{(s)} \mid X^{(\ell)} = x) - \lambda C_s\}.$$

Stopping further search with $s = 0$ yields always a continuation value of 0. An unbiased observation of $Q_{\lambda,s}^{(\ell)}(x)$ at value $x = X^{(\ell),i}$ is thus given by replacing the expectation of the payoff with the path payoff, so that the right-hand side of the equation above is approximated at the point $x = X^{(\ell),i}$ by

$$Q_{\lambda,s}^{(\ell)}(x) = 2^{s-\ell}\{\hat{P}_\lambda^{(s),i} - \lambda C_s\} + \varepsilon,$$

where $\varepsilon$ has zero expectation, $E(\varepsilon) = 0$, if the strategy fitted at layers larger than $\ell$ is optimal. If it is only approximately optimal, there will be a, hopefully small, bias in $\varepsilon$. It follows that, for each layer $\ell$ and action $s \in \{\ell+1, \ldots, G\}$, there are $m$ independent samples available to fit the continuation value. As the continuation value is monotonically increasing in its argument, one can find a fit $\hat{Q}_{\lambda,s}^{(\ell)}$ as the monotone increasing function that minimizes

$$(18) \qquad \hat{Q}_{\lambda,s}^{(\ell)} = \operatorname*{arg\,min}_{f:\,f\uparrow} 2^{s-\ell} \sum_{i=1}^{m} (\tilde{P}_{\lambda}^{(s),i} - \lambda C_s - f(X^{(s),i}))^2.$$

In summary, the algorithm proceeds in analogy to the scheme outlined in Section 2.4.4 as follows:

I **Simulate Monte Carlo samples.** Simulate $m$ random paths in the tree. For each of the $m$ paths, simulate the distribution of $X$ over this path. The resulting Monte Carlo samples are, for each layer $\ell = 1, \ldots, G$, given by $X^{(\ell),i}$ with $i = 1, \ldots, m$.

II **Compute continuation values.**

  (1) **Initialize leaf nodes.** Set current generation $\ell$ to the highest generation $G$. The payoff under the optimal strategy is

$$\hat{P}_{\lambda}^{(G)}(x) = 1\{x \geq q\}.$$

  (2) **Compute continuation value.** Move one generation down, $\ell \leftarrow \ell - 1$. The continuation value $Q_{\lambda,s}^{(\ell)}(x)$ is computed for all $s \in \{\ell+1, \ldots, G\}$ separately according to (18). The payoff $\hat{P}_{\lambda}^{(\ell),i}$ for the $i$th sample is then determined for $s \neq 0$ by

$$\hat{P}_{\lambda}^{(\ell),i} = 2^{s-\ell}\{\hat{P}^{(s),i} - \lambda C_s\}$$

  and 0 if $s = 0$. The decision $s$ on the r.h.s. above is given by

$$s = \operatorname*{arg\,max}_{s'} \hat{Q}_{\lambda,s'}^{(\ell)}(X^{(\ell),i})$$

  and by $s = 0$ if the maximum of the argument on the r.h.s. over all values $s'$ is negative.

  (4) **Loop.** If $\ell > 1$, go to Step 2. Else stop.

If the number of sample paths $m$ is converging to infinity, the fitted continuation values, and hence the fitted strategy, will converge to the optimal continuation values and the optimal strategy. We refrain from giving a proof, as it is standard DP, only slightly adapted to the situation at hand. The crucial step is the replacement of simulation over entire trees by simulation along paths.

2.5.4. *Choosing the alternative.* Note that the optimal strategy depends on the underlying distribution $X$. If the distribution of the original observations $Z$ is known, then the distribution of $X$ over the entire tree is in general also known (in the sense that one can simulate from it). Knowing the distribution of $Z$ requires, however, that we know how many false null hypotheses with $\theta_j \in \Theta_A$ we expect among all $n$ observations and what the distribution of $Z$ is under the alternative.

The alternative is often unknown. One approach is to choose the alternative distribution for $\theta_j \in \Theta_A$ such that any false null hypothesis is just barely detectable under naive search. Still, this approach would not determine the precise shape of the distribution of $Z$ under the alternative. We have had quite good experience with an alternative approach. Simulating under the global null hypothesis, we merely try to detect all strong (and rare) peaks in the test statistics. We might be losing a bit of power in this way, as the strategy is possibly not optimal. Yet, this turned out to be a negligible effect in practice. The advantage is that no assumption about the distribution under the alternative has to be made.

The chosen approach helps also with another practical issue. If the threshold $q$ is rather large, there are just a tiny fraction of sample paths which exceed the threshold in the final layer. The number of sample paths required (to achieve a reasonable fit) is then very large. With simulation under the global null, we can simply adjust the threshold $q$ to a very high quantile of the null distribution. As a rule of thumb, if we try to reduce the computational burden to a fraction $\beta$ of the naive search, one could aim for a threshold $q$ that corresponds to the $1 - \beta$ quantile of the null distribution of all $Z_j$, $j = 1, \ldots, n$ values. If every exceedance of this threshold is detected by the scheme (and nothing else), the computational burden would be in the same order of magnitude as the fraction $\beta$ of the cost of the naive search, as every detection in the final layer is preceded by a path of observations in the previous layers. This rule of thumb seems to work well in practice.

**3. Numerical example.** We apply the hierarchical search scheme to the data of pulsar PSR B1706–44. In the, roughly, 14 day long observation of this pulsar, 1072 photons with energies above 200 MeV were observed by EGRET. It is known for this pulsar that the true frequency at the beginning of the observations is roughly 9.761175993 Hz. The pulsar spins down with a drift of $-8.827879 \cdot 10^{-12} \mathrm{s}^{-2}$. Over the entire observational time, the pulsar went through roughly $10^8$ cycles. In a typical blind search for pulsars, a frequency range from 1–40 Hz would have to be searched and a drift range from 0 down to $-5 \cdot 10^{-11} \mathrm{s}^{-2}$. Searching over a sufficiently fine grid would require in the range of $10^9$ frequency-drift pairs to be searched.

We emphasize that we use PSR B1706–44 merely as a prototype, providing example data for a blind search. This pulsar was detected first using radio observations. It is only a known gamma-ray pulsar because a targeted (not blind) search at the known period finds a significant gamma-ray signal. Although thousands of radio pulsars are known, only a handful pulse at high energies (X rays and gamma rays), where astronomers count individual photons, and all of these but one were detected in a targeted search. Detection of rare high-energy pulsars would be very exciting and novel.

Here, we fit the hierarchical scheme with five successive layers. The final layer computes the test statistics $F(\omega, \dot{\omega})$ in (1) on a grid with 3 observation points between frequencies which are spaced $1/T$ apart and 3 drift values between $1/T^2$-spaced drift values. For each of the layers $\ell = 1, \ldots, 4$ beneath, the times series is divided into $2^{5-\ell}$ blocks of equal length and the test statistic $F^{5-\ell}(\omega, \dot{\omega})$ in (2) is calculated, for each layer $\ell = 1, \ldots, 5$ on a grid with spacing of length $\delta\omega^\ell = 2^{5-\ell}/(3T)$ in frequency space and $\delta\dot{\omega}^\ell = 2^{2(5-\ell)}/(9T^2)$ in drift space, effectively giving an oversampling of factor 3 in both the frequency and drift direction. The search tree is thus an 8-ary tree. Each node can be represented in terms of the associated frequency $\omega$ and drift $\dot{\omega}$. The children of a node $(\omega, \dot{\omega})$ in layer $\ell$ (the children being in layer $\ell + 1$) correspond to the frequency drift combinations

$$(\omega + \eta_\omega \delta\omega^{\ell+1}, \dot{\omega} + \eta_{\dot{\omega}} \delta\dot{\omega}^{\ell+1})$$

for all 8 possible combinations of $\eta_\omega \in \{-1/2, 1/2\}$ and $\eta_{\dot{\omega}} \in \{-3/2, -1/2, 1/2, 3/2\}$.

To fit the optimal strategy, we simulate $5 \cdot 10^5$ random paths under the global null hypothesis. As motivated above, the threshold $q$ is chosen as the 0.999 quantile of the distribution of $F(\omega, \dot{\omega})$, which are the observations in the final layer. One crucial aspect is the choice of $\lambda$. Each value of $\lambda$ gives a different tradeoff between computational cost and detection power. Here, we adjusted $\lambda$ such that the computational cost of searching is below 0.1% of the cost of the naive search (0.08% to be precise). The cost clearly depends on the underlying distribution of $X$. If there are many false null hypotheses in the final layer, computation is slowed down as more time is spent on examining those peaks. We measured the cost under a global null hypothesis. The reasoning is that the vast majority of computational resources are spent on regions in frequency-drift space which are well approximated by this global null assumption.

Figure 5 shows the outcome of the search for pulsar PSR B1706–44. We restrict ourselves to a very small region around the true frequency and drift. It can be seen how the search narrows down regions of interest. In the first layer, all values are examined on a coarse grid. The regions of interest are refined on finer and finer grids in successive layers. At the final layer, the peak at the true values of frequency and drift is among the observed values and would thus be found in a blind search. The question arises clearly how easy it is to find the peak. The test statistic $F$ at the true frequency and drift is approximately equal to a p-value of $1.13 \cdot 10^{-7}$ (this has been calculated using a chi-squared approximation for $F$ with 2 degrees of freedom). While this may sound like a large peak (and hence an easy rejection), it has to be taken into account that one is searching over hundreds of millions of frequency-drift pairs. To adjust for multiplicity, one can use a Bonferroni correction. This is not very conservative if one is looking just for a single rejection to be made. Clearly, there is dependence between the test statistics at nearby frequencies and drift values, but we could take, as a lower bound for the multiplicity correction, the number of gridpoints in the frequency-drift scheme, divided by 9 (as 3 gridpoints are placed between frequencies spaced $1/T$ apart and frequencies $1/T$ apart are approximately independent). At level 0.05, the peak of PSR B1706–44 would only be in the rejection region if up to half a million tests were made. In the current problem, the multiplicity correction would have to be in the tens or hundreds of millions. Hence, the peak is found with the search scheme even though the signal is very weak and is below the rejection threshold (which is a property of the signal itself, not of our hierarchical search).

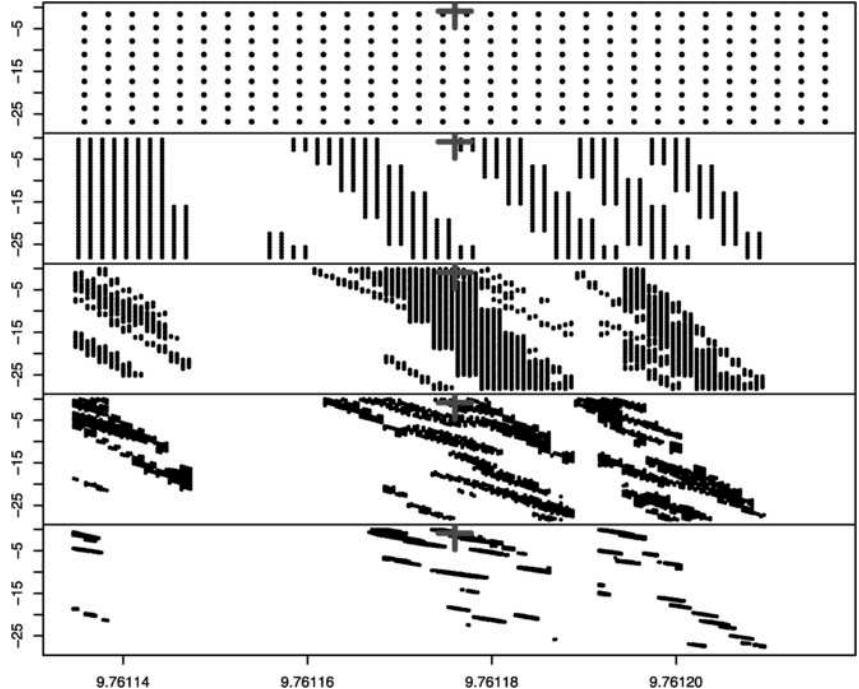

Fig. 5.  *Outcome of the testing scheme for pulsar PSR B1706–44 in the vicinity of the true frequency and drift (marked with a cross). From left to right, the dots indicate all observed frequency-drift pairs in layers 1 to 5 (with number of blocks decreasing from $2^4 = 16$ to $2^0 = 1$). At the highest level, all pairs are examined on a coarse grid, while lower levels search for interesting frequency-drift pairs on an increasingly finer grid. The true frequency-drift combination is found at the finest resolution in the lowest layer. The shown area corresponds only to a tiny fraction of the frequency-drift space to be searched. The search activity in other regions is much less.*

It is important to emphasize that detection in a blind search, as proposed here, requires (a) that the correct frequency-drift pair is identified by the hierarchical search strategy and (b) that the signal at the correct frequency-drift is above a threshold needed to claim detection. Note that success in (b) is a property of the signal and not related to our hierarchical search, but stronger signals make correct identification in step (a) surely easier. For PSR B1706–44, the signal is not strong enough to reject the null hypothesis and (b) is not satisfied. Despite this weak signal, the correct frequency-drift (as known from radio signals) is identified in step (a) by our hierarchical search. This is very reassuring and makes it plausible that we would also identify the correct frequency-drift combinations of pulsars that are of interest in a true blind search, as they would need to have a signal strong enough to pass the threshold in step (b). Their frequency-drift combination is thus even more likely to be identified by our hierarchical search.

Next, the power of the detection scheme is examined for simulated data. All parameters are the same as in the example of pulsar PSR B1706–44, including the fact that 1072 photons are observed from a time-varying source. However, the density of photon arrival phases is now controlled. We assume a simple sinusoidal density. For a given frequency $\omega$ and drift $\dot{\omega}$, the density for an observed phase $\phi = \omega t + \dot{\omega} t^2/2$ is proportional to $1 + \theta \sin(2\pi\phi)$, for some value $0 < \theta < 1$. Larger values of $\theta$ make detection easier, while $\theta = 0$ corresponds to a aperiodic source, the null hypothesis. Note that the density above can be modulated by any function $c(t)$ without changing the result noticeably, as long as $c(t)$ is nearly constant over time-scales of length $1/\omega$.

We choose, for each simulation, randomly a frequency $\omega$ and drift $\dot{\omega}$ in the search range of 1 to 40 Hz and 0 to $-5 \cdot 10^{-11} s^{-2}$, respectively. On the one hand, we calculated the power to detect any periodicity for the "naive" blind search, as a function of $\theta$. Every test statistic above the $1 - \alpha/n$-quantile of the null distribution of $F$ is rejected. The level $\alpha$ was set to 0.05. The multiplicity correction $n$ was set to $10^9$. Rejections require in this case roughly $20\sigma$ events.

The hierarchical search was implemented with various values of the trade-off parameter $\lambda$, using $10^5$ random samples paths under the global null hypothesis as training samples. The threshold $q$ for training of the hierarchical scheme was chosen lower than the actual required threshold for rejection (as motivated above). It was set as in the example above to the 0.999 quantile of the null distribution of $F(\omega, \dot{\omega})$ under the global null. The implicit goal of the hierarchical search is thus to identify any $8\sigma$ event in the final layer. Any such event might not be rejected, as rejections require $20\sigma$ events. Tuning the scheme toward higher thresholds might increase its efficiency but would also require many more sample paths for the fitting stage. The fitting of the continuation values took merely a few seconds on a standard CPU and this cost is hence negligible compared to the cost of the search.

The aim is to see how the tradeoff parameter $\lambda$ influences both the computational cost and the power, which is here the probability that the "true" frequency-drift pair $(\omega, \dot{\omega})$ is rejected. For the "true" frequency-drift pair to be rejected, (a) the relevant test statistic has got to be above the final rejection threshold and (b) the test statistic has got to be observed (not been dropped by the search scheme before it reaches the final layer). The power of detection is now a function of $\lambda$, as the chance of observing the relevant test statistic is a function of $\lambda$. We count every rejection as successful if it is within $1/T$ of the "true" frequency and within $1/T^2$ of the "true" drift. For $\lambda \to 0$, we obtain the same power as for the "naive" blind search, as every hypothesis in the final layer is examined. For larger values of $\lambda$, the power of detection starts to decrease, while the computational cost decreases. Figure 6 shows some results. We use four values $\theta = 0.24, 0.26, 0.29, 0.34$. With these values, the power for the "naive" blind search is, approximately, increasing from 0.1245 for $\theta = 0.24$ over power of 0.2381 and 0.4920 to a power of 0.8717 for $\theta = 0.34$. The computational cost and detection power is calculated for 1000 simulations for each value of $\theta$ and $\lambda$. As can be seen from Figure 6, the computational cost can be lowered by more than three orders of magnitude, while still retaining more than 90% of the power of the blind search.

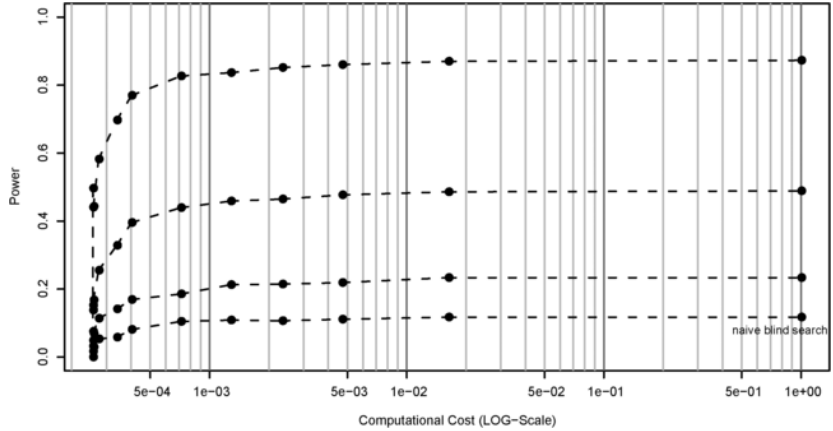

FIG. 6. *The tradeoff between computational cost and detection power. Four different signals strengths are used. The computational cost of the naive search (looking at every hypotheses in the final layer) is set to 1. The power starts to deteriorate substantially once the computational cost is reduced by more than three orders of magnitude compared with the naive blind search.*

**Acknowledgments.** We would like to thank a referee, the Associate Editor and the Editor for many insightful and very helpful comments on an earlier version.

N. Meinshausen
Department of Statistics
University of Oxford
1 South Parks Road
Oxford OX1 3TG
United Kingdom
E-mail: meinshausen@stats.ox.ac.uk

P. Bickel
J. Rice
Department of Statistics
University of California—Berkeley
367 Evans Hall
Berkeley, California 94720
USA
E-mail: bickel@stat.berkeley.edu
        rice@stat.berkeley.edu